\documentclass[11pt,a4paper]{article}
\usepackage{jcappub}

\title{Constraints on $\Lambda(t)$CDM models as holographic and agegraphic dark energy with the observational Hubble parameter data}
\author[a]{Zhong-Xu Zhai}
\author[b,c]{Tong-Jie Zhang}
\author[a]{and Wen-Biao Liu}

\affiliation[a]{Department of Physics, Institute of Theoretical
Physics, Beijing Normal University, Beijing, 100875, China}
\affiliation[b]{Department of Astronomy, Beijing Normal University,
Beijing, 100875, China}
\affiliation[c]{Center for High Energy Physics, Peking University, Beijing, 100871, China}

\emailAdd{zzx@mail.bnu.edu.cn} \emailAdd{tjzhang@bnu.edu.cn}
\emailAdd{wbliu@bnu.edu.cn}

\abstract{ The newly released observational $H(z)$ data (OHD) is
used to constrain $\Lambda(t)$CDM models as holographic and
agegraphic dark energy. By the use of the length scale and time
scale as the IR cut-off including Hubble horizon (HH), future event
horizon (FEH), age of the universe (AU), and conformal time (CT), we
achieve four different $\Lambda(t)$CDM models which can describe the
present cosmological acceleration respectively. In order to get a
comparison between such $\Lambda(t)$CDM models and standard
$\Lambda$CDM model, we use the information criteria (IC), $Om(z)$
diagnostic, and statefinder diagnostic to measure the deviations.
Furthermore, by simulating a larger Hubble parameter data sample in
the redshift range of $0.1<z<2.0$, we get the improved constraints
and more sufficient comparison. We show that OHD is not only able to
play almost the same role in constraining cosmological parameters as
SNe Ia does but also provides the effective measurement of the
deviation of the DE models from standard $\Lambda$CDM model. In the
holographic and agegraphic scenarios, the results indicate that the
FEH is more preferable than HH scenario. However, both two time
scenarios show better approximations to $\Lambda$CDM model than the
length scenarios. }

\begin{document}
\maketitle

 \flushbottom


\section{Introduction} \label{sec:introduction}

In the past more than ten years, the accelerating expansion of the
universe has obtained a lot of attention since the observations of
type Ia supernovae \cite{SNe..Riess,SNe..Hicken}. Later on, the
discoveries of CMB \cite{CMB..Spergel,CMB..Komatsu} and BAO
\cite{BAO..Eisenstein,BAO..Percival} observations also got the same
results. In order to give a reasonable explanation to this scenario,
a great variety of attempts have been done. These theoretical
explanations can be classified into two categories. The first is the
dark energy which involve the introduction of exotic matter sources.
The second is the modified gravity which relates to the changes of
the geometry of the spacetime
\cite{dark..energy..review..Tsujikawa}. Among these approaches,
$\Lambda$CDM, as the standard model, is considered to be the
simplest and most natural one which shows great consistence with the
observational data. In this model, the cosmological constant
$\Lambda$ is considered to be the dark energy component of the
universe. However, this model is also challenged by the fine-tuning
problem and the coincidence problem. In order to solve or alleviate
these problems, several methods are proposed and discussed
\cite{cons..prob..S..Weinberg,cons..pro..A. Vilenkin,cons..pro..J.
Garriga,cons..pro..G. Caldera-Cabral}. One possible way is to treat
$\Lambda$ no longer a real constant but a time-dependent variable,
namely the $\Lambda(t)$CDM
\cite{Lt..Peeble,Lt..Ozer,Lt..Sahni,Lt..Wang,Lt..S. Basilakos}. In
the assumption, the form of $\Lambda(t)$ is considered to be related
to the cosmological length scale or time scale \cite{Lt..scale..R.
Bousso}
\begin{equation}
\Lambda(t)=\frac{3}{r_{\Lambda}^{2}(t)}=\frac{3}{t_{\Lambda}^{2}},
\end{equation}
where the speed of light is set as $c=1$, $r_{\Lambda}$ and
$t_{\Lambda}$ are the length scale and time scale which can
introduce a nonzero and time-varying $\Lambda(t)$. Thus, it is
natural to connect the $\Lambda(t)$CDM model with the holographic
and agegraphic dark energy models which are proposed by the
holographic principle \cite{holographic..principle..Cohen}. In these
scenarios, the dark energy takes the form
\begin{equation}
  \rho_{\Lambda}=3c^{2}M_{P}^{2}\frac{1}{L^{2}},
\end{equation}
where $c$ is a numerical factor, $M_{P}^{2}=1/8\pi G$ is the reduced
Planck mass, and $L$ is the size which represents the IR cut-off or
the UV-cutoff. Recently, several kinds of cosmological scales are
considered to be the role of IR cut-off. These scenarios are
summarized in \cite{lengthscale..timescale..Chen} including the
Hubble horizon, the particle horizon, the future event horizon as
the length scale and the age of the universe, the conformal time as
the time scale
\cite{hubble..horizon..Xu,future..horizon..Li,age..constant..Xu,time..constant..Zhang..Zhang,time..constant..3Zhang}.
Therefore, there are five kinds of $\Lambda(t)$CDM models so far.
The different choices of these scales give various evolutions of
$\Lambda(t)$ and different dynamical behavior of the universe. Thus
it is necessary to compare the theoretical analysis with the
cosmological observations. This direction has been explored in
\cite{lengthscale..timescale..Chen} which uses the recently compiled
``Union2 Compilation" of Type Ia supernovae (SNe Ia) data.

As another kind of the cosmological observation, the Hubble
parameter $H$ which is directly related to the expansion history of
the universe $H=\dot{a}/a$ has been tested in several cosmological
models
\cite{Hz..constraint..Yi,Hz..constraint..Samushia,Hz..constraint..Wei,Hz..constraint..Wu,
Hz..constraint..Lazkoz,Hz..constraint..Kurek,Hz..constraint..Cao1,Hz..constraint..Cao2,Hz..constraint..Zhai,Hz..constraint..Zhang,Hz..constraint..Zhangxin}.
Different from the distance scale measurements such as SNe Ia, CMB
and BAO, there is no integral in calculating $H(z)$. This feature
makes the Hubble parameter preserve the information related to the
cosmological evolution. Thus investigate the observational $H(z)$
data is very rewarding. The previous works showed that the
observational Hubble data (OHD) can give a good supplement to the
cosmic observations and can also be used as an alternative to the
SNe Ia data which is considered as the "standard
candle"\cite{Hubble..parameter..LinHui,fR..Hubble.vs.redshift..Carvalho}.
So, whether the OHD can give a compatible constraints with the SNe
Ia on the $\Lambda(t)$CDM model should be focused on. However,
updated and expanded several times, the amount of available $H(z)$
data is still scarce compared with SNe Ia luminosity distance data
\cite{Hubble..expand..Jimenez,Hubble..expand..Simon,Hubble..expand..Stern}.
One way to avoid this embarrassment and alleviate the inaccuracy of
lack of data points is to expand the data set of OHD by simulation.
In our calculation, we will apply the method proposed in
\cite{Hubble..parameter..MaCong} which shows that as many as 64
independent OHD data points can match the parameter constraining
power of SNe Ia by comparing the median figures of merit.

Except that, it is also useful to employ the information criteria to
assess these $\Lambda(t)$CDM models since their origin of
holographic and agegraphic are similar with each other. In this
paper, we use the AIC, BIC and AIC$_{c}$ as model selection criteria
to distinguish different models from the the statistical view.
Furthermore, we also adopt the $Om(z)$ and statefinder diagnostics
to evaluate the dark energy models and measure their derivation from
the standard $\Lambda$CDM model. The corresponding analysis and
discussion are presented in section \ref{IC}

Our paper is organized as follows. In section \ref{basic}, we present
the basic formulas of $\Lambda(t)$CDM models with an interaction
term between dark energy and dark matter. In section \ref{hubbledata},
the observational Hubble data is described. The constraints on the
$\Lambda(t)$CDM models as holographic and agegraphic dark energy are
presented in section \ref{constraints}. In section \ref{IC}, we compare and
measure the deviation of such $\Lambda(t)$CDM models from
$\Lambda$CDM model. The simulated Hubble parameter data sample and
its corresponding constraints are shown in section \ref{simulation}. Our
discussion and conclusion are summarized in section \ref{discussions}.

\section{$\Lambda(t)$CDM model in the flat FRW universe}\label{basic}

In the frame of Einstein gravity, considering the flat FRW metric
\begin{equation}
  ds^{2}=-dt^{2}+a(t)^{2}(dr^{2}+r^{2}d\theta^{2}+r^{2}\sin^{2}\theta d\varphi^{2}),
\end{equation}
with $a(t)$ standing for the scale factor, we can get the Friedmann equation
\begin{equation}\label{Eq:Friedman}
H^{2}=\frac{1}{3M_{P}^{2}}(\rho_{m}+\rho_{\Lambda}),
\end{equation}
where $H$ is Hubble parameter, $\rho_{m}$ and $\rho_{\Lambda}$ are the energy densities of the matter and vacuum (dark energy) respectively.
By the use of the dimensionless density parameters $\Omega_{m}=\rho_{m}/3M_{P}^{2}H^{2}$ and $\Omega_{\Lambda}=\rho_{\Lambda}/3M_{P}^{2}H^{2}$,
this equation can be rewritten as
\begin{equation}
\Omega_{m}+\Omega_{\Lambda}=1.
\end{equation}
In our scenario, we suppose that the energy can exchange between matter and vacuum energy through the interaction term $Q$, thus the local energy-momentum
conservation law can yield
\begin{equation}\label{Eq:exchange}
  \dot{\rho}_{m}+3H(1+\omega_{m})\rho_{m}=Q, \qquad  \dot{\rho}_{\Lambda}+3H(1+\omega_{\Lambda})\rho_{\Lambda}=-Q,
\end{equation}
where "." means the derivative with respect to the cosmic time $t$ and the sign in front of $Q$ will decide the direction of the energy flux.
$\omega_{m}=0$ and $\omega_{\Lambda}=-1$ are the equation of state (EOS) for the matter and vacuum energy respectively. With the definitions of the
effective EOS for both the energy term
\begin{equation}
  \omega_{m}^{eff}=\omega_{m}-\frac{Q}{3H\rho_{m}}, \qquad \omega_{\Lambda}^{eff}=\omega_{\Lambda}+\frac{Q}{3H\rho_{\Lambda}},
\end{equation}
eq.(\ref{Eq:exchange}) can be rewritten as
\begin{equation}\label{Eq:EOS}
    \dot{\rho}_{m}+3H(1+\omega_{m}^{eff})\rho_{m}=0, \qquad  \dot{\rho}_{\Lambda}+3H(1+\omega_{\Lambda}^{eff})\rho_{\Lambda}=0.
\end{equation}
Substituting this equation to eq.(\ref{Eq:Friedman}), we get the evolution of Hubble parameter $H=H_{0}E(z)$ where the expansion rate $E(z)$ can be
expressed as
\begin{equation}\label{Eq:expansionrate}
  E^{2}(z)=\Omega_{m0}\exp({3\int_{0}^{z}\frac{1+\omega_{m}^{eff}}{1+z'}dz'})+\Omega_{\Lambda0}\exp(3{\int_{0}^{z}\frac{1+\omega_{\Lambda}^{eff}}{1+z'}dz'}),
\end{equation}
where the subscript "0" denotes the present value of a quantity.


\section{The observational Hubble parameter data}\label{hubbledata}

The measurement of Hubble parameter $H(z)$ is increasingly becoming important in cosmological constraints.
As a direct measurement related to the expansion history of the universe, it can be derived from the derivation
of redshift $z$ with respect to the cosmic time $t$ \cite{Hubble..parameter..derived..Jimenez}
\begin{equation}
  H(z)=-\frac{1}{1+z}\frac{dz}{dt}.
\end{equation}
From the observation of galaxy ages, Jimenez et al. demonstrated the feasibility of the method by applying it to a
$z\sim0$ sample \cite{Hubble..expand..Jimenez}. Simon et al. further derived a set of OHD including 8 data points with a redshift range up to $z=1.75$
from the relative ages of passively evolving galaxies \cite{Hubble..expand..Simon}. Additionally, Stern et al. obtained an expanded data set and constrained the
cosmological parameters \cite{Hubble..expand..Stern}. In our paper, we will focus on these data listed above \cite{Hz..constraint..Zhang,Hubble..parameter..MaCong}and constrain the $\Lambda(t)$CDM as holographic and agegraphic dark energy models.

By the use of $\chi^{2}$ statistics, one can get the best-fitting values of the parameters and its corresponding confident regions
\begin{equation}
  \chi^{2}(H_{0},z,p)=\sum_{i}(H_{th,i}(H_{0},z_{i},p)-H_{obs,i}(z_{i}))^{2}/\sigma_{i}^{2})
\end{equation}
where $p$ stands for the parameter vector of each dark energy models and $H_{0}=73.8\pm2.4$
is the prior which is a 3\% precision measurement suggested in \cite{Hubble..new..prior..Riess}.

\begin{figure}[htb]
\begin{center}
\includegraphics[width=0.45\textwidth]{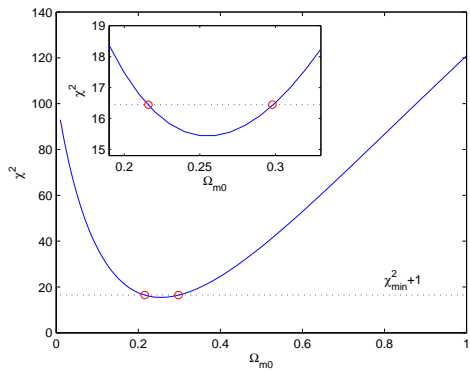}
\end{center}
\caption{The $\chi^{2}$ evolution with respect to $\Omega_{m0}$ for $\Lambda$CDM model. The circles stand for the 68.3\% confidence interval and
the insert gives zoom-in views around the $1\sigma$ confident level.}
\label{fig:LCDMconstraints}
\end{figure}

In the calculations, we choose the standard $\Lambda$CDM model as the fiducial model which gives the expression of $H(z)$ as
\begin{equation}
  H_{fid}(z)=H_{0}E(z)=H_{0}\sqrt{\Omega_{m0}(1+z)^{3}+\Omega_{\Lambda}}.
\end{equation}
The constraints on this fiducial model by OHD are $\chi^{2}_{\min}=15.4478$, $\Omega_{m0}=0.26_{-0.04}^{+0.04}$ and $\Omega_{\Lambda}=0.73_{-0.04}^{+0.04}$.
And the $\chi^{2}\sim \Omega_{m}$ relation is plotted in figure \ref{fig:LCDMconstraints}
It can be seen that the best fit values are consistent with the 7-year WMAP \cite{CMB..Komatsu}, the BAO \cite{BAO..Percival} and SNe Ia \cite{SNe..Hicken}.
In the next section, we will constrain the holographic and
agegraphic dark energy models and compare their results with the $\Lambda$CDM model.

\section{Constraints on the holographic and agegraphic dark energy models}\label{constraints}

\subsection{Hubble horizon as the IR cut-off} \label{sec:hubblehorizon}

The Hubble horizon is defined as
\begin{equation}\label{Eq:Hubblehorizon}
  d_{H}=\frac{1}{H}.
\end{equation}
It provides an estimate of the distance that light can travel while the universe expands appreciably. When the Hubble horizon is chosen as the IR cut-off
of the holographic dark energy, we can get
\begin{equation}
\Lambda(t)=3c^{2}H^{2}(t), \quad \rho_{\Lambda}=3c^{2}M_{P}^{2}H^{2},
\end{equation}
where $c$ is a constant instead of the speed of light and satisfying \cite{hubble..horizon..Xu}
\begin{equation}
  c^{2}<1.
\end{equation}
Thus one can get
\begin{equation}
  \Omega_{\Lambda}=\frac{\rho_{\Lambda}}{3M_{P}^{2}H^{2}}=c^{2}.
\end{equation}
Consequently, the expansion rate of this scenario is
\begin{equation}
  E(z)=(1+z)^{3(1-c^{2})}.
\end{equation}
So, the choice of Hubble horizon as the IR cut-off (HH) presents a one-parameter cosmological model.

\begin{figure}[htb]
\begin{center}
$\begin{array}{cc}
\includegraphics[width=0.5\textwidth,height=0.4\textwidth]{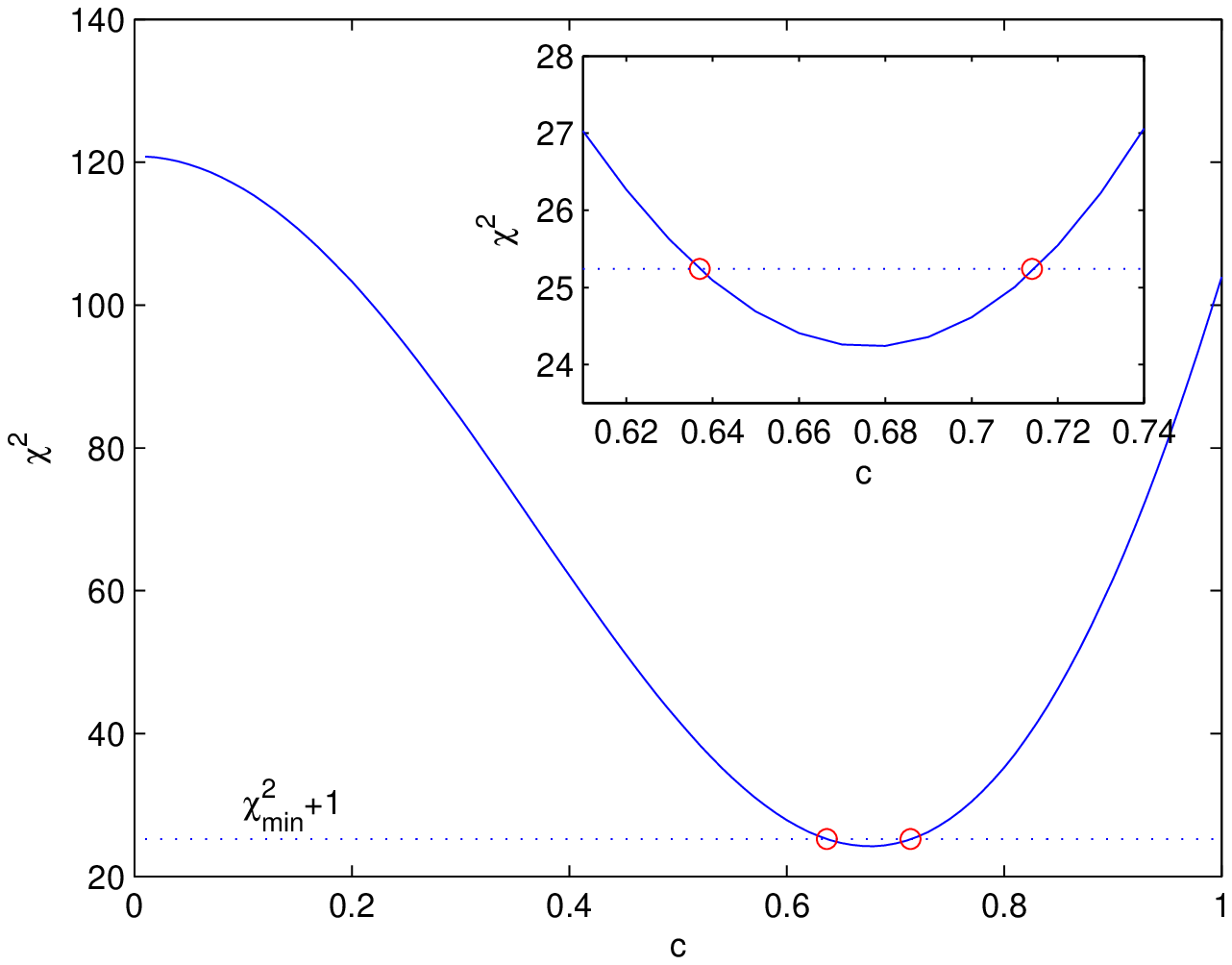} &
\includegraphics[width=0.5\textwidth,height=0.4\textwidth]{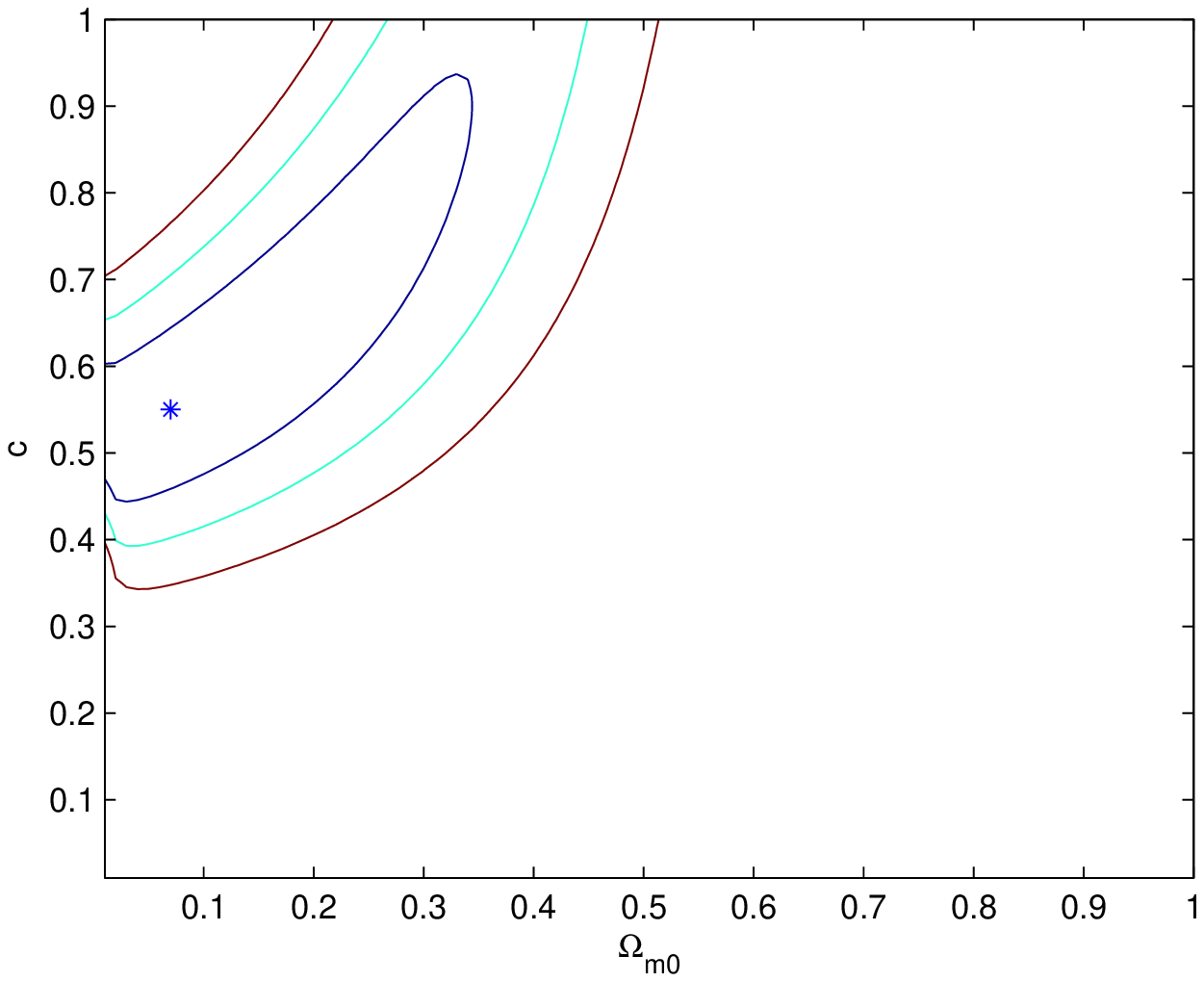}
\end{array}$
\end{center}
\caption{$Left$: The $\chi^{2}$ evolution with respect to $c$ for HH scenario. The circles stand for the $1\sigma$ confident interval and the insert gives zoom-in views around the 68.3\% confidence level.
$Right$: The 68.3\%, 95.4\% and 99.7\% confident regions of FEH scenario from inner to outer. The best-fit values is marked by the star.}
\label{fig:lengthOHD}
\end{figure}

The application of $\chi^{2}$ statistic and OHD give a constraint of the parameter $c=0.68_{-0.04}^{+0.03}$ on the 68.7\% confidence level, i.e.
$\Omega_{m0}=0.54_{-0.04}^{+0.05}$, a bit larger than the observations suggested \cite{CMB..Komatsu,BAO..Percival,SNe..Hicken}. The evolution of $\chi^{2}$ with $c$ is plotted in FIG.\ref{fig:lengthOHD} and
$\chi^{2}_{\min}=24.2393$.

\subsection{Particle horizon as the IR cut-off}

The particle horizon is defined as
\begin{equation}
  d_{p}=a(t)\int_{0}^{t}\frac{dt'}{a},
\end{equation}
which is the distance that light have traveled since the beginning of the universe.
The calculation in \cite{lengthscale..timescale..Chen} give the effective EOS of vacuum energy
\begin{equation}
  \omega_{\Lambda}^{eff}=-\frac{1}{3}+\frac{2}{3c}\sqrt{\Omega_{\Lambda}},
\end{equation}
where $c$ describe the deviation from the de Sitter universe. It can be seen that the choice of particle horizon can not lead to
the present cosmic acceleration because $\omega_{\Lambda}^{eff}>-1/3$. So there is no need to constrain its undetermined parameters.

\subsection{Future event horizon as the IR cut-off}

The future event horizon is defined as
\begin{equation}
  d_{E}=a\int_{t}^{\infty}\frac{dt'}{a},
\end{equation}
which is the distance that light will be able to travel in the future. By the use of the same method as the Hubble horizon,
we can get
\begin{equation}
  \Lambda(t)=3c^{2}/d_{E}^{2}, \quad \rho_{\Lambda}=3c^{2}M_{P}^{2}/d_{E}^{2},
\end{equation}
where $c$ is taken account to fill the deviation from the de Sitter universe\cite{future..horizon..Li}. Several steps after, one can get the evolution of $\Omega_{\Lambda}$ with
respect to the redshift $z$
\begin{equation}
  \frac{d\Omega_{\Lambda}}{dz}=-\Omega_{\Lambda}(1-3\Omega_{\Lambda}+\frac{2}{c}\sqrt{\Omega_{\Lambda}})(1+z)^{-1}.
\end{equation}
Furthermore, the effective EOS for both matter and vacuum energy can also be achieved
\begin{equation}
\omega_{\Lambda}^{eff}=-\frac{1}{3}-\frac{2}{3c}\sqrt{\Omega_{\Lambda}},\quad \omega_{m}^{eff}=-\frac{2}{3}(1-\frac{\sqrt{\Omega_{\Lambda}}}{c})\frac{\Omega_{\Lambda}}{1-\Omega_{\Lambda}}.
\end{equation}
Substituting these to eq.(\ref{Eq:expansionrate}), we can get the evolution of the expansion rate $E(z)$.

Apparently, the choice of future event horizon (FEH) derives a two-parameters cosmological model which can lead to the accelerating phase. When performing the
$\chi^{2}$ statistics for the OHD to constrain the parameters $(\Omega_{m0},c)$, we get $\Omega_{m0}=0.07_{-0.07}^{+0.27}$ and $c=0.55_{-0.10}^{+0.39}$ at the 68.7\% confidence level.
This best results show a universe with fewer matter comparing with the observations. However, at the $1\sigma$ confident level, the value of $\Omega_{m0}$ is consistent
with the observations.
The probability contours on the $(\Omega_{m0},c)$ plane is plotted in figure \ref{fig:lengthOHD} (the right panel)
with $\chi^{2}_{\min}=14.7359$.

\subsection{Age of the universe as the IR cut-off}

The age of the universe is defined as
\begin{equation}
  t_{\Lambda}=\int_{0}^{t}dt'.
\end{equation}
Under this scenario, we obtains
\begin{equation}
  \Lambda(t)=3c^{2}/t_{\Lambda}^{2}, \quad \rho_{\Lambda}=3c^{2}M_{P}^{2}/t_{\Lambda}^{2},
\end{equation}
where $c$ is a constant to fill the derivation from the de Sitter universe \cite{age..constant..Xu}. After several steps of calculation, we can get the evolution of $\Omega_{\Lambda}$ satisfying
\begin{equation}\label{Eq:ageomega}
\frac{d\Omega_{\Lambda}}{dz}=-\Omega_{\Lambda}(3-3\Omega_{\Lambda}-\frac{2}{c}\sqrt{\Omega_{\Lambda}})(1+z)^{-1},
\end{equation}
and two EOSs for matter and vacuum energy
\begin{equation}\label{Eq:ageEOS}
\omega_{\Lambda}^{eff}=-1+\frac{2}{3c}\sqrt{\Omega_{\Lambda}}, \quad \omega_{m}^{eff}=\frac{2\sqrt{\Omega_{\Lambda}}}{3c}\frac{\Omega_{\Lambda}}{1-\Omega_{\Lambda}}.
\end{equation}
Substituting both eq.(\ref{Eq:ageomega}) and eq.(\ref{Eq:ageEOS}) into eq.(\ref{Eq:expansionrate}), one can determine the expansion rate of this model.

\begin{figure}[htb]
\begin{center}
$\begin{array}{cc}
\includegraphics[width=0.5\textwidth,height=0.4\textwidth]{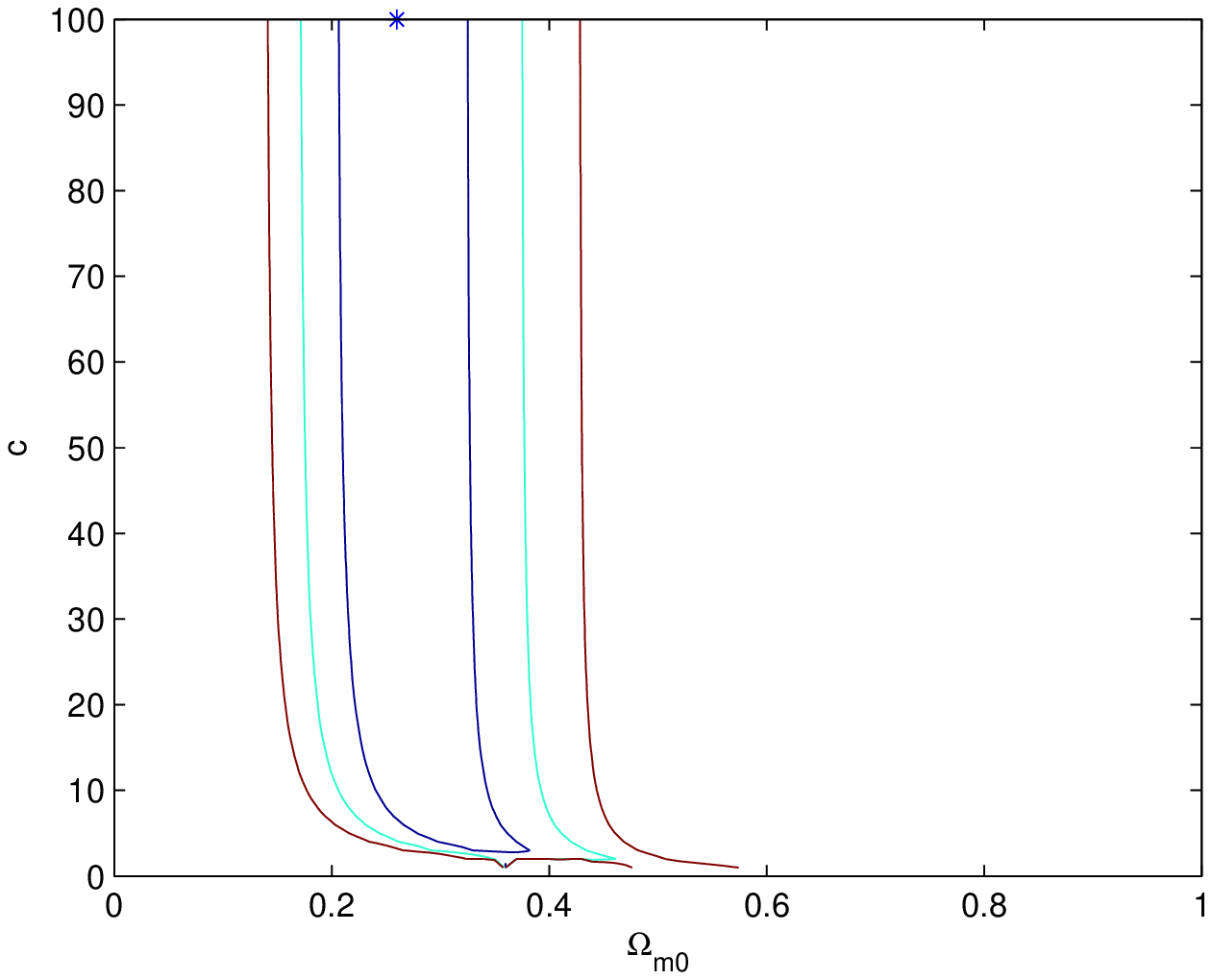} &
\includegraphics[width=0.5\textwidth,height=0.4\textwidth]{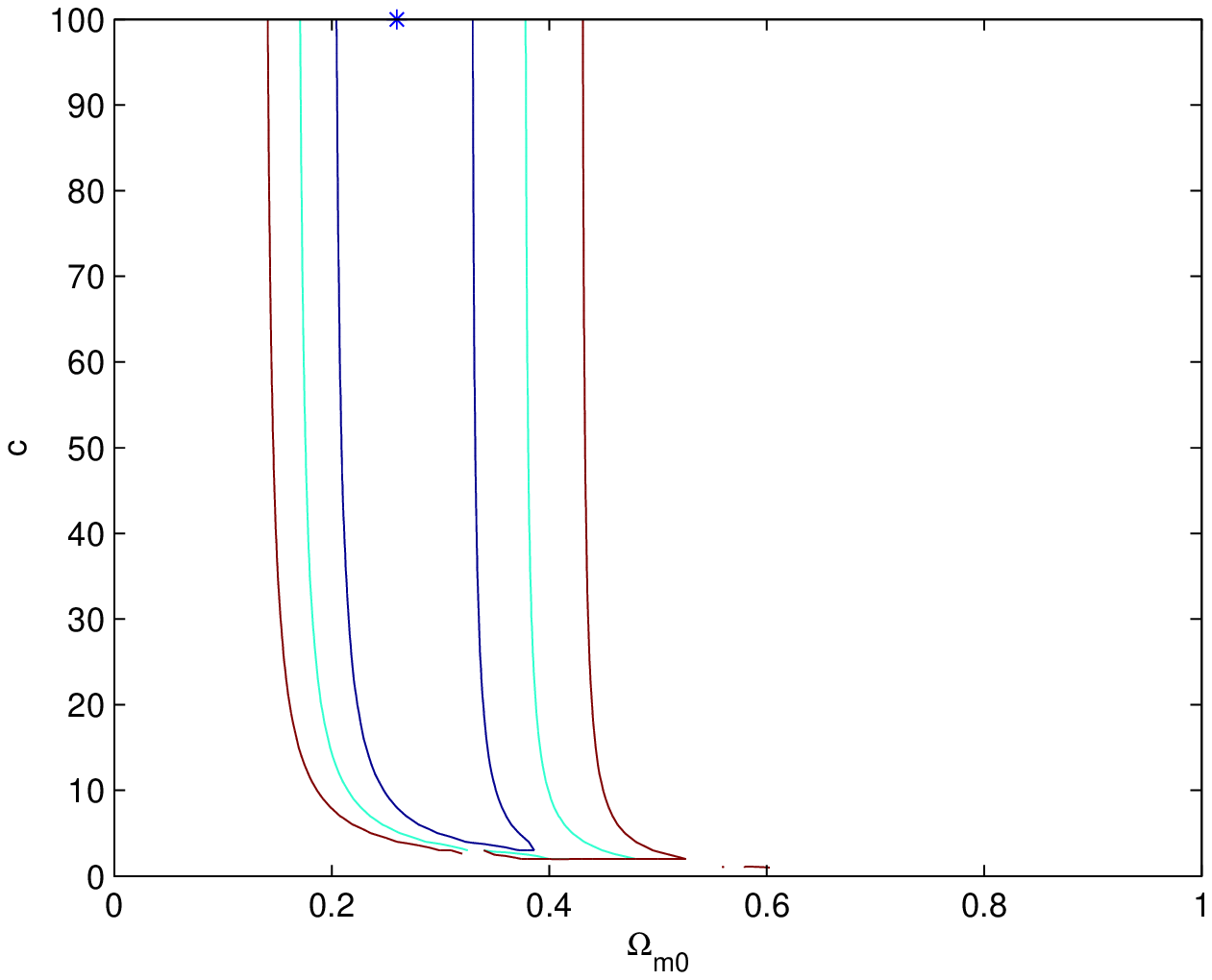}
\end{array}$
\end{center}
\caption{The 68.3\%, 95.4\% and 99.7\% confident regions from inner to outer in the $(\Omega_{m0}-c)$ plane, while the star means the best-fit values.
$Left$: AU scenario; $Right$: CT scenario.}
\label{fig:timeOHD}
\end{figure}

The same as the previous section, the choice of age of the universe as the IR cut-off (AU) also gives a two-parameter cosmological model. Using the OHD and
$\chi^{2}$ statistics, we get the best-fit results of the model parameters: $\Omega_{m0}=0.26_{-0.06}^{+0.18}$ and $c\in(3.0,\infty)$ at the 68.3\% confidence level.
It should be noticed that the OHD can not give a finite best-fit value and upper limit of $c$ while the constraint just shows a lower limit of $1\sigma$.
In figure \ref{fig:timeOHD}, we plot the confident regions on the $(\Omega_{m0},c)$ plane and we set $c=100$ as the cut-off.

\subsection{Conformal time as the IR cut-off}

The conformal time is defined as
\begin{equation}
  \eta_{\Lambda}=\int_{0}^{t}\frac{dt'}{a},
\end{equation}
which is the total comoving distance light could travel \cite{conformal..time..Myung}. In this case, we can get
\begin{equation}
   \Lambda(t)=3c^{2}/\eta_{\Lambda}^{2}, \quad \rho_{\Lambda}=3c^{2}M_{P}^{2}/\eta_{\Lambda}^{2}.
\end{equation}
By the use of the same method as presented in the previous section, one can obtain the differential relation of $\Omega_{\Lambda}$ with respect to redshift $z$ and
the effective EOSs for both matter and vacuum energy
\begin{equation}\label{Eq:conformalomega}
  \frac{d\Omega_{\Lambda}}{dz}=-\Omega_{\Lambda}\big[\frac{3(1-\Omega_{\Lambda})}{1+z}-\frac{2}{c}\sqrt{\Omega_{\Lambda}}\big],
\end{equation}
\begin{equation}\label{Eq:conformalEOS}
  \omega_{\Lambda}^{eff}=-1+\frac{2}{3c}(1+z)\sqrt{\Omega_{\Lambda}}, \quad \omega_{m}^{eff}=-\frac{2(1+z)\sqrt{\Omega_{\Lambda}}}{3c}\frac{\Omega_{\Lambda}}{1-\Omega_{\Lambda}}.
\end{equation}
The corresponding expansion rate of this model can be obtained by substituting this equation into eq.(\ref{Eq:expansionrate}).

Similar to the previous models, the choice of conformal time as the IR cut-off (CT) also derives a two-parameter cosmological model. The constraint of OHD
gives the best-fit results $\Omega_{m0}=0.26_{-0.06}^{+0.18}$ and $c\in(3.0,\infty)$ which is almost the same as the scenario of age of the universe.
The absences of the best-fit value and upper limit of $c$ also exist in this model.
The only difference is $\chi^{2}_{\min}=15.5016$ a little bit larger than the one obtained in the age of the universe scenario which is $\chi^{2}_{\min}=15.4952$.
figure \ref{fig:timeOHD} shows the confident region on the $(\Omega_{m0},c)$ plane while setting $c=100$ as the cut-off.


\section{Model Identification}\label{IC}
\subsection{Information criteria}
It is widely known that $\chi^{2}$ statistics is effective to find the best-fit values of parameters in a model.
However it can not be used to decide which model is the best one among various models sufficiently. In our work, we have chosen
the standard $\Lambda$CDM model as the fiducial model, so it is necessary to assess the holographic and agegraphic
dark energy models with the fiducial model. The information criteria (IC) is an effective way to assess different models.
In the following, we will use the BIC and AIC as model criteria. In this criteria frame,
models which give lower values of BIC and AIC will be preferred\cite{AIC..BIC..Li,AIC..BIC..Liddle,AIC..BIC..Szydlowski,AIC..BIC..Godlowski,AIC..BIC..Biesiada}.

The BIC is given by\cite{BIC..definition}
\begin{equation}
  BIC=-2\ln\mathcal {L}_{max}+k\ln N
\end{equation}
where $\mathcal {L}_{max}$ is the maximum likelihood and can be calculated as $-2\ln\mathcal {L}_{max}=\chi^{2}_{min}$, $k$ is the number of
the model parameters and $N$ is the number of the data points. The difference in $BIC$ is denoted as $\Delta BIC_{j}=BIC_{j}-BIC_{min}$ where $j$
stands for different models. The model that can give a smaller $\Delta BIC$ will be considered closer to the fiducial one.

The AIC is defined as\cite{AIC..definition}
\begin{equation}
AIC=-2\ln\mathcal {L}_{max}+2k,
\end{equation}
where $\mathcal {L}_{max}$ and $k$ have the same meanings as in BIC. Similarly, the one which gives the smallest value of AIC will be considered as the best model.
The difference of AIC from the fiducial one is $\Delta AIC_{j}=AIC_{j}-AIC_{min}$ which describes the deviation from the fiducial model.
Additionally, for the small data sample, there is a corrected AIC version which is defined as\cite{AICc..definition}
\begin{equation}
  AIC_{c}=AIC+\frac{2k(k-1)}{N-k-1}.
\end{equation}
This formula is useful when $N/k\leq40$. In our calculation, there is only 13 data points in OHD, so $AIC_{c}$ is suitable.

Our results of information criteria are listed in Table. I. We can see that all three information criteria results show that the standard $\Lambda$CDM
model is the best one and this is consistent with the assumption that setting $\Lambda$CDM as the fiducial model at the beginning of our calculation.
 Although the OHD constraint on FEH scenario give the smallest value of $\chi_{min}^{2}$, the information criteria (both BIC and AIC) show a
punishment because of its complexity. Except that, the large value of information criteria in HH shows a great deviation from the fiducial model.
From the previous constraints, we see that the best-fit results obtained in AU and CT scenarios are very similar. Furthermore, their values of information criteria
both located around 2. This situation shows a positive evidence that the AU and FT scenarios are close to the fiducial model.
\begin{table}[!h]\label{TB:1}
\tabcolsep 0pt
\vspace*{-12pt}
\begin{center}
\def\temptablewidth{0.8\textwidth}
{\rule{\temptablewidth}{1pt}}
\begin{tabular*}{\temptablewidth}{@{\extracolsep{\fill}}ccccccccc}
  & $\chi_{min}^{2}$ & $\Delta\chi_{min}^{2}$ & BIC & $\Delta$BIC & AIC & $\Delta$AIC & AIC$_{c}$ & $\Delta$AIC$_{c}$ \\   \hline
    $\Lambda$CDM   & 15.4478 & 0       & 18.0127 & 0      & 17.4478 & 0      & 17.4478 & 0  \\
            HH     & 24.2393 & 8.7915  & 26.8042 & 8.7915 & 26.2393 & 8.7915 & 26.2393 & 8.7915\\
           FEH     & 14.7359 & -0.7119 & 19.8658 & 1.8530 & 18.7359 & 1.2881 & 19.1359 & 1.6881 \\
            AU     & 15.4952 & 0.0474  & 20.6251 & 2.6123 & 19.4952 & 2.0474 & 19.8952 & 2.4474 \\
            CT     & 15.5016 & 0.0538  & 20.6315 & 2.6187 & 19.5016 & 2.0538 & 19.9016 & 2.4538
       \end{tabular*}
       {\rule{\temptablewidth}{1pt}}
       \end{center}
       \caption{Summary of the information criteria analyze}
       \end{table}
\subsection{The $Om(z)$ diagnostic}
\begin{figure}[htb]
\begin{center}
\includegraphics[width=0.55\textwidth]{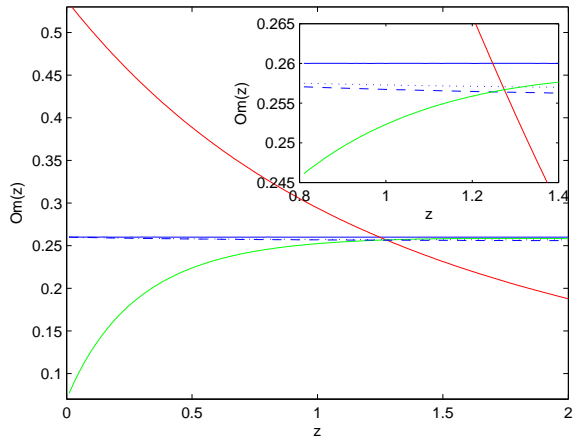}
\end{center}
\caption{The $Om(z)$ diagnostic as a function of redshift $z$. The solid blue curve shows the standard $\Lambda$CDM model and the red and green one
belong to HH and FEH respectively. The AU and CT models are shown by the dotted and dashed line.
The insert gives zoom-in views near the intersections.}
\label{fig:Omz}
\end{figure}

In order to get a more transparent comparison between the fiducial model and $\Lambda(t)$CDM models, we use the $Om(z)$ diagnostic which is given by \cite{Omz..definition}
\begin{equation}
  Om(z)=\frac{E^{2}(z)-1}{(1+z)^{3}-1}.
\end{equation}
Apparently, $Om(z)=\Omega_{m0}$ for $\Lambda$CDM. Thus this diagnostic is demonstrated to be useful to distinguish $\Lambda$CDM from other DE models.
Except that, $Om(z)$ relies only on the knowledge of Hubble parameter. This makes the errors in the reconstruction of $Om$ are bound to be small.
In figure \ref{fig:Omz}, we plot the evolutions of $Om(z)$ for each model. We can see that the HH and FEH scenarios behaves different from $\Lambda$CDM model apparently.
However, the two agegraphic dark energy models show great consistence with $\Lambda$CDM where $Om(z)$ remains constant. It is natural because the best-fit
value of $c$ in the two agegraphic dark energy models takes the limit $c\rightarrow\infty$. According to eq.(\ref{Eq:ageEOS}) and eq.(\ref{Eq:conformalEOS}),
these two model both have $\omega_{m}^{eff}\rightarrow0$ and $\omega_{\Lambda}^{eff}\rightarrow-1$. This limit makes them return to the standard $\Lambda$CDM model and get to
a situation that the interaction term can be neglected.

\subsection{Statefinder diagnostic}

The statefinder diagnostic was first introduced in \cite{statefinder..defi..Sahni} which is capable to discriminate various dark energy models. This parameter pair$\{s,r\}$
is given as
\begin{equation}\label{Eq:state}
  r=\frac{\dddot{a}}{aH^{3}},\qquad s=\frac{r-1}{3(q-1/2)},
\end{equation}
where $q$ is the deceleration parameter. Obviously, this statefinder pair is constructed directly from the spacetime metric, i.e the second and
third derivatives of the scale factor $a$. Since the trajectories in the $s-r$ plane belong to different cosmological models show qualitatively
different behaviors, the statefinder diagnostic can be seen as a powerful tool to distinguish different cosmological models.
So far, many works have been done by means of the statefinder diagnostic analysis, such as the standard $\Lambda$CDM model, the quintessence with and without interacting models,
the holographic dark energy models, the phantom model, and the tachyon model
\cite{statefinder..app..Zhang3,statefinder..app..Malekjani,statefinder..app..Alam,statefinder..app..Zimdahl,statefinder..app..Zhang1,statefinder..app..Zhang2,statefinder..app..Chang,statefinder..app..Shao}.

Firstly, from the definition of the deceleration factor $q=-(\ddot{a}a)/\dot{a}^{2}$ and eq.(\ref{Eq:Friedman}), one can get
\begin{equation}
  q=\frac{1}{2}-\frac{3\Omega_{\Lambda}}{2}.
\end{equation}
Because the parameter $r$ in eq.(\ref{Eq:state}) can be rewritten as
\begin{equation}\label{Eq:r}
  r=\frac{\ddot{H}}{H^{3}}-3q-2,
\end{equation}
substituting the relation
\begin{equation}
  \frac{\ddot{H}}{H^{3}}=\frac{9}{2}(-\omega_{\Lambda}^{eff}+1-2\Omega_{\Lambda}),
\end{equation}
to eq.(\ref{Eq:r}), we can obtain
\begin{equation}
  r=1-\frac{9}{2}\Omega_{\Lambda}(1+\omega_{\Lambda}^{eff}),
\end{equation}
\begin{equation}
  s=1+\omega_{\Lambda}^{eff}.
\end{equation}

In the flat FRW universe, the standard $\Lambda$CDM model corresponds to a fixed point
\begin{equation}
  \{s,r\}_{\Lambda\text{CDM}}=\{0,1\}.
\end{equation}
This value is independent on the parameter of $\Lambda$CDM model and cosmic time $t$(or equivalently redshift $z$).
The statefinder diagnostic applied to the above $\Lambda(t)$CDM models as holographic and agegraphic dark energy derives several
formulas of $\{s,r\}$, these results are summarized in Table. II.

\begin{table}[!h]
\tabcolsep 5mm
\vspace*{12pt}
\begin{center}
\begin{tabular}{|r|c|c|}\hline
model & s & r \\ \hline
HH & 1-$c^{2}$ & $c^{2}$\\ \hline
FEH & $\frac{2}{3}-\frac{2}{3c}\sqrt{\Omega_{\Lambda}}$ & $1-3\Omega_{\Lambda}+\frac{3}{c}\Omega_{\Lambda}^{3/2}$\\ \hline
AU & $\frac{2}{3c}\sqrt{\Omega_{\Lambda}}$ & $1-\frac{3}{c}\Omega_{\Lambda}^{3/2}$\\ \hline
CT & $\frac{2}{3c}(1+z)\sqrt{\Omega_{\Lambda}}$ & $1-\frac{3}{c}(1+z)\Omega_{\Lambda}^{3/2}$ \\ \hline
\end{tabular}
\end{center}
\caption{The statefinder pair for the $\Lambda(t)$CDM models}
\end{table}

\begin{figure}[htb]
\begin{center}
$\begin{array}{cc}
\includegraphics[width=0.5\textwidth,height=0.4\textwidth]{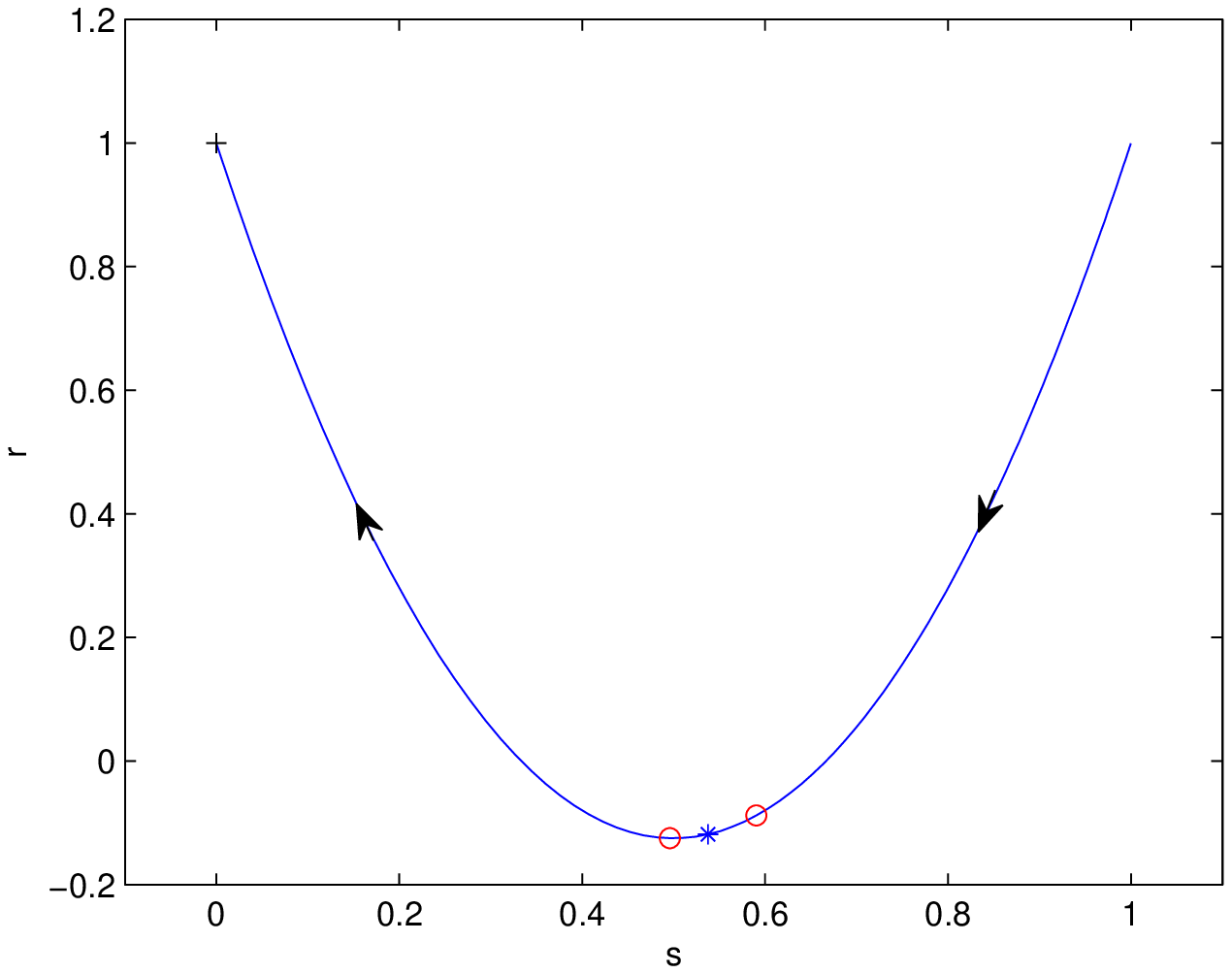} &
\includegraphics[width=0.5\textwidth,height=0.4\textwidth]{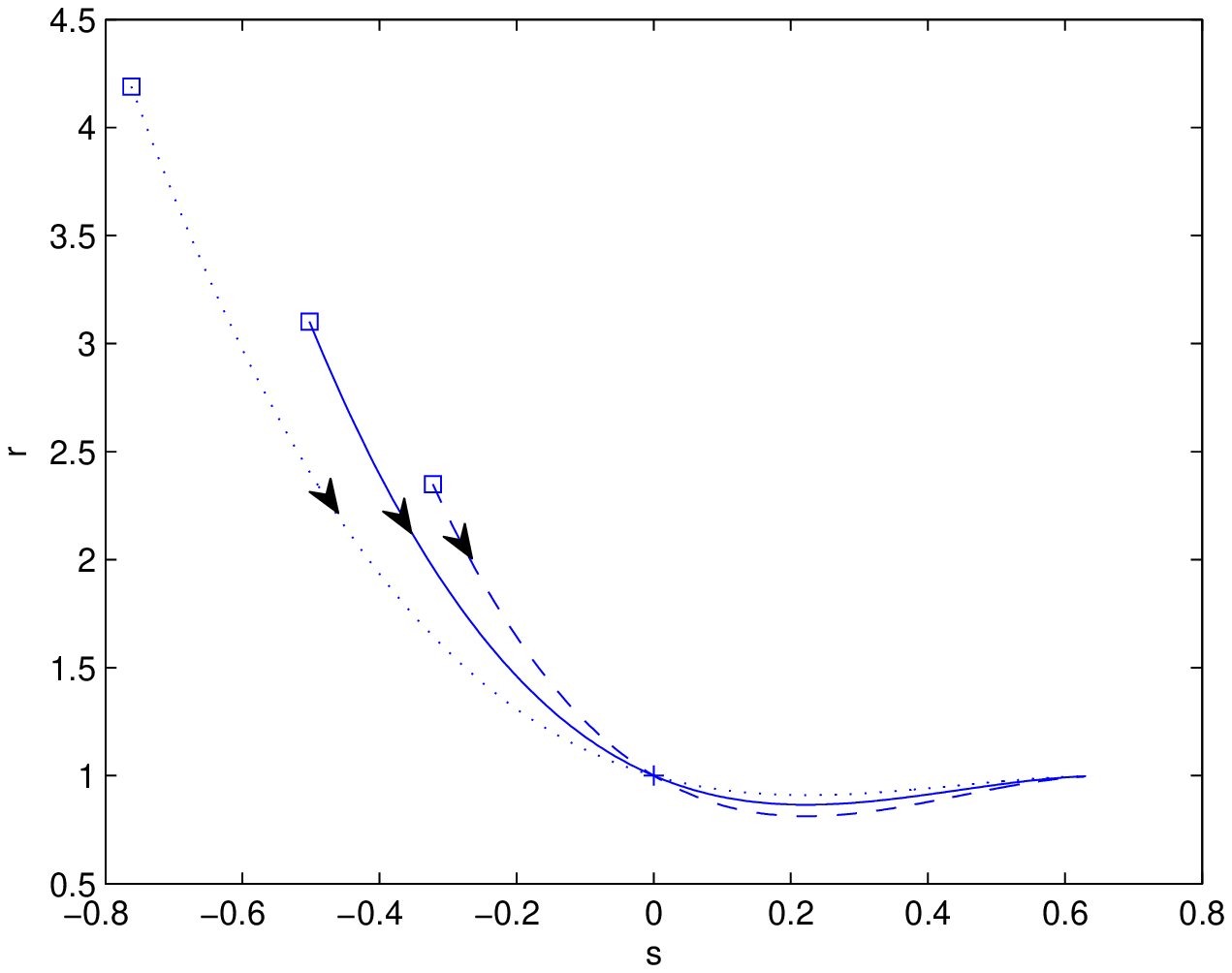}
\end{array}$
\end{center}
\caption{$Left$: The evolutionary trajectory of the statefinder in the $s-r$ plane for the HH scenario with respect to the parameter $c$. The star and
red circles stand for the situations of $1\sigma$ interval of $c=0.68_{-0.04}^{+0.03}$. The arrows show the varying direction of $c=0\rightarrow1$.
The standard $\Lambda$CDM model is denoted by the cross.
$Right$: The evolutionary trajectory of the statefinder in the $s-r$ plane for the FEH scenario with different $(\Omega_{m0},c)$ values:
$(\Omega_{m0},c)=(0.07,0.45)$(dotted line), $(\Omega_{m0},c)=(0.07,0.55)$(solid line) and $(\Omega_{m0},c)=(0.07,0.94)$(dashed line).
The arrows show the redshift-varying direction $z=0\rightarrow\infty$ with squares standing for the present values.
The standard $\Lambda$CDM model is denoted by the cross.}
\label{fig:state1}
\end{figure}

\begin{figure}[htb]
\begin{center}
$\begin{array}{cc}
\includegraphics[width=0.5\textwidth,height=0.4\textwidth]{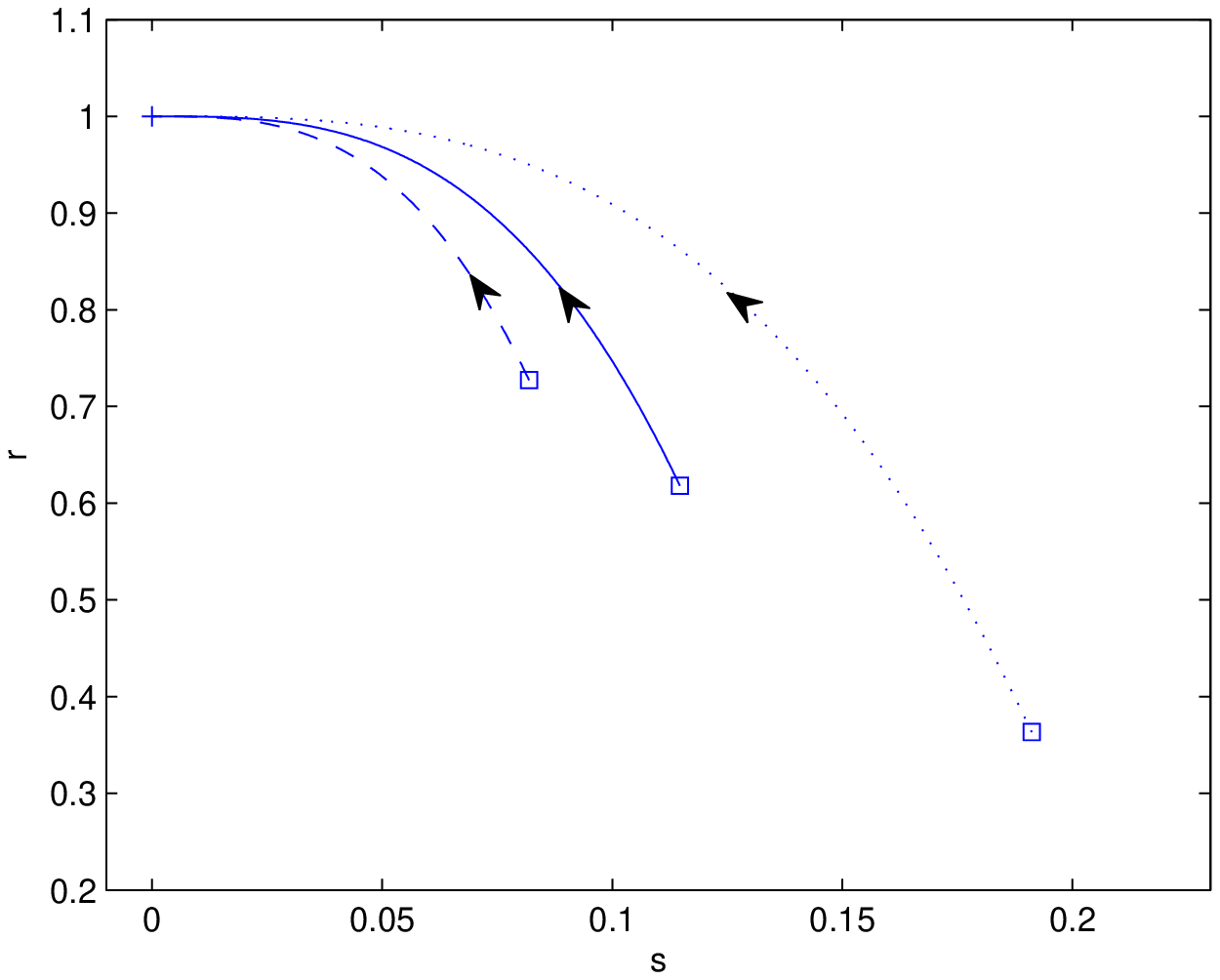} &
\includegraphics[width=0.5\textwidth,height=0.4\textwidth]{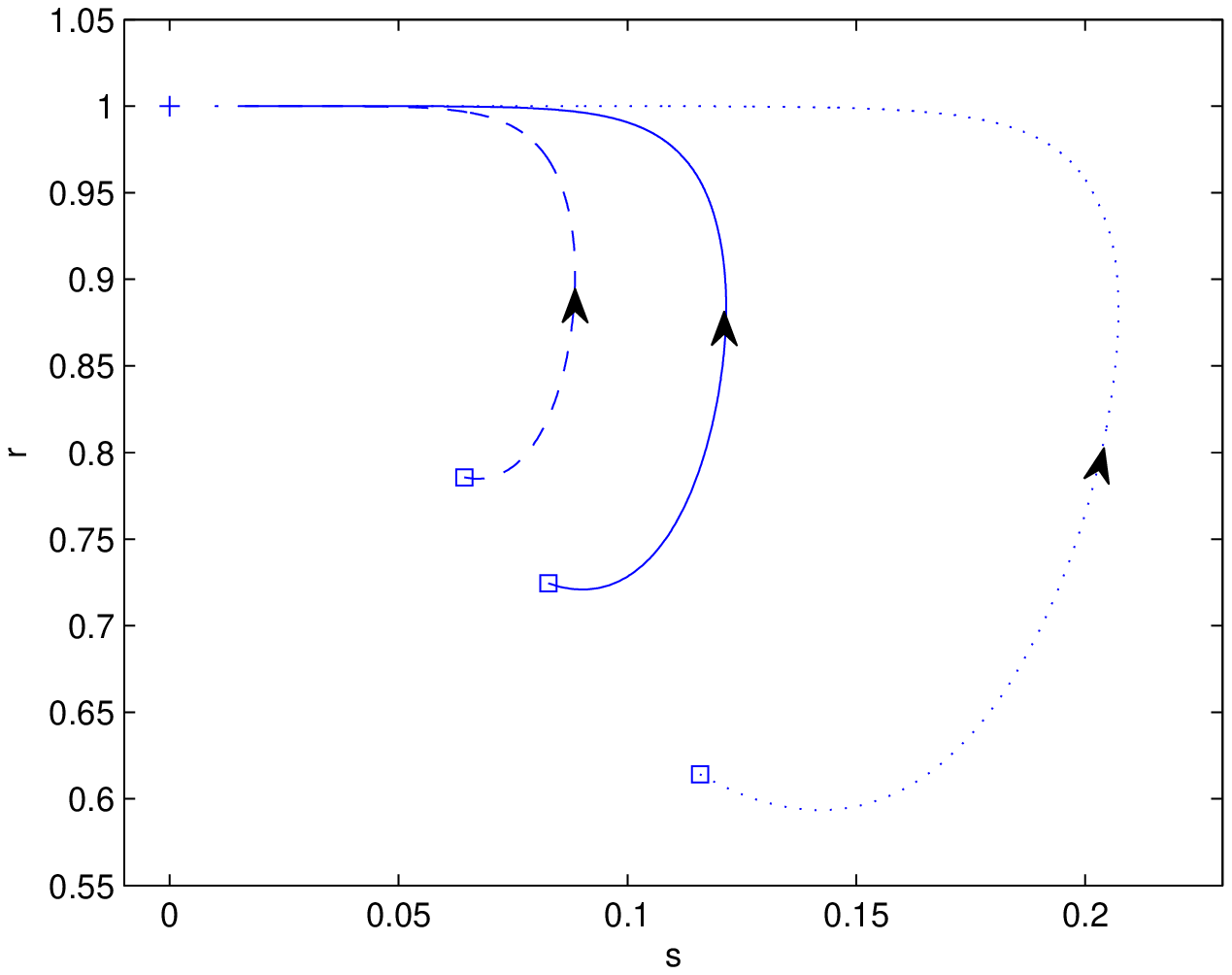}
\end{array}$
\end{center}
\caption{$Left$: The evolutionary trajectory of the statefinder in the $s-r$ plane for the AU scenario with different $(\Omega_{m0},c)$ values:
$(\Omega_{m0},c)=(0.26,3)$(dotted line), $(\Omega_{m0},c)=(0.26,5)$(solid line) and $(\Omega_{m0},c)=(0.26,7)$(dashed line).
The arrows show the redshift-varying direction $z=0\rightarrow\infty$ with squares standing for the present values.
The standard $\Lambda$CDM model is denoted by the cross.
$Right$: The evolutionary trajectory of the statefinder in the $s-r$ plane for the CT scenario with different $(\Omega_{m0},c)$ values:
$(\Omega_{m0},c)=(0.26,5)$(dotted line), $(\Omega_{m0},c)=(0.26,7)$(solid line) and $(\Omega_{m0},c)=(0.26,9)$(dashed line).
The arrows show the redshift-varying direction $z=0\rightarrow\infty$ with squares standing for the present values.
The standard $\Lambda$CDM model is denoted by the cross.}
\label{fig:state2}
\end{figure}

In figure \ref{fig:state1} and \ref{fig:state2}, we plot the evolutionary trajectories of statefinder for the $\Lambda(t)$CDM as holographic and agegraphic
dark energy models in the $s-r$ plane. The same as the standard $\Lambda$CDM  model, the statefinder pair $\{s,r\}$ in the HH model is also a fixed point
but is dependent on the parameter $c$. By the use of the best fit results from OHD constraints, we obtain $\{s,r\}=\{0.5376,-0.1186\}$ in HH scenario which shows a great derivation
from $\Lambda$CDM model as the previous analyze. The arrows in the curve show that the universe full of vacuum energy $\Omega_{\Lambda}=1$ in the HH scenario
can give the same statefinder pair as $\Lambda$CDM model $\{s,r\}=\{0,1\}$.

Different from the $\Lambda$CDM model and HH scenario, the statefinder pairs $\{s,r\}$ of the three two-parameter models FEH, AU and CT are all parameter-dependent
and relying on the redshift $z$. Applying the constraining results of OHD, we present several trajectories of statefinder for these models with different $c$ values.
From the right panel of figure \ref{fig:state1}, we get that there is a quintessence-like to phantom-like transition of the dark energy with the cosmic time increasing.
Naturally, at the moment of transition, the statefinder $\{s,r\}=\{0,1\}$ returns to $\Lambda$CDM model. From the numerical solution of $\Omega_{\Lambda}$ in \cite{lengthscale..timescale..Chen},
it can be seen that $\Omega_{\Lambda}$ approaches to 0 as the redshift $z$ increases. This result derives that the statefinder $\{s,r\}$ of FEH will reach to a limit
$\{s,r\}=\{2/3,1\}$ at the very beginning of the universe.

The evolutionary trajectories of statefinder for the two age cosmological constant models AU and CT are quite different (see figure \ref{fig:state2}). This diagnostic distinguishes
AU and CT helpfully. However, we should note that the values of $c$ used here are just located in the $1\sigma$ confident level $(3,\infty)$. When applying
$c=\infty$, the statefinder pairs of AU and CT in the $s-r$ plane will approach to the fixed point (0,1) as the standard $\Lambda$CDM model which are still indistinguishable.
Certainly, with the increasing of $c$, the range of the statefinder curve decreases and finally degenerates to a point. Except that, no matter what finite value
of $c$ is chosen, the $\{s,r\}$ will reach to the $\Lambda$CDM situation at the very beginning of the universe.


\section{Generation of simulated datasets and constraints}\label{simulation}

\begin{figure}[htb]
\begin{center}
\includegraphics[width=0.55\textwidth]{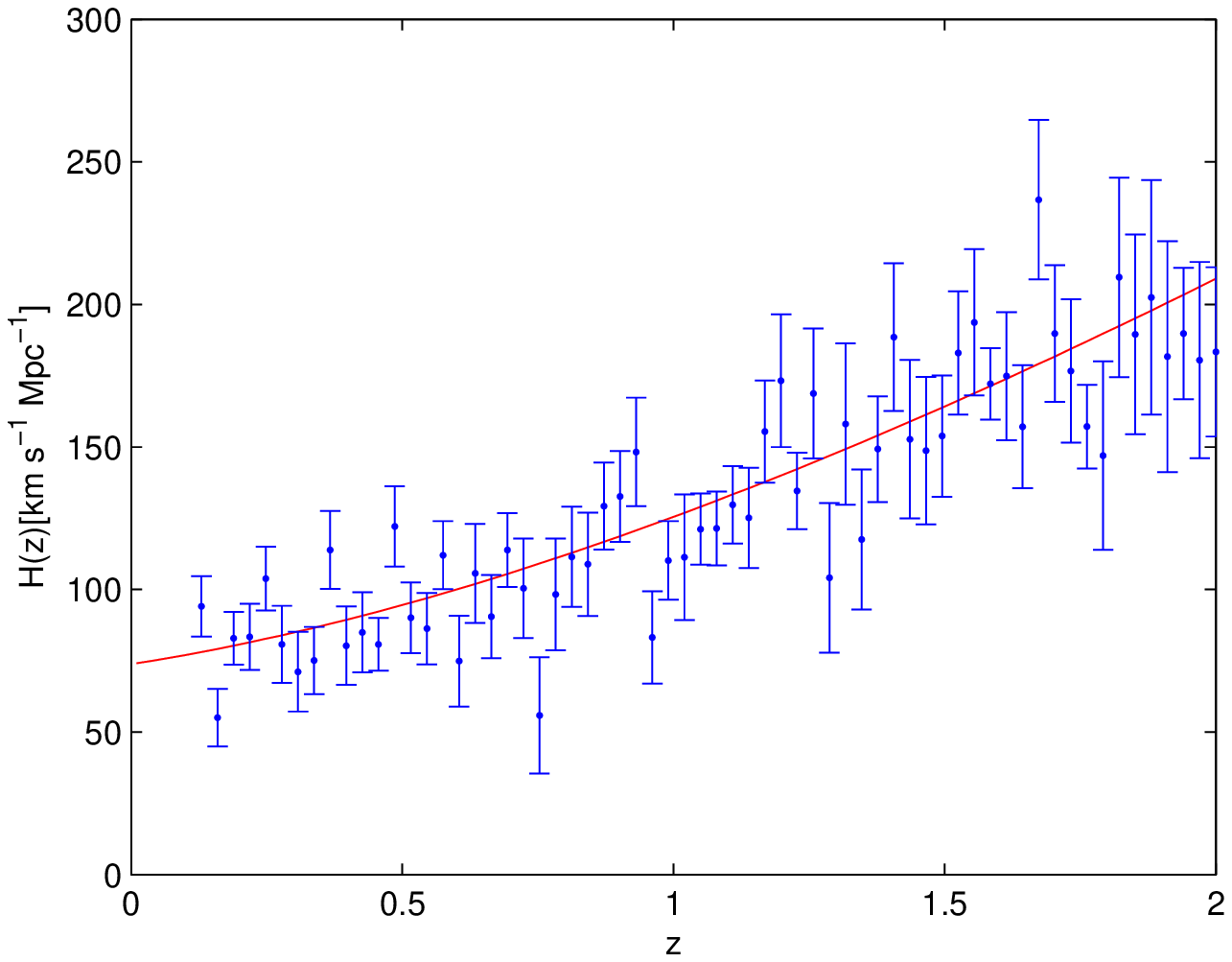}
\end{center}
\caption{Snapshot of s simulated $H(z)$ dataset with $1\sigma$ error bar in the redshift range of $0.1<z<2.0$. The red curve stands for the theoretical
evolution of $H(z)$ in the fiducial model.}
\label{fig:SHD}
\end{figure}

So far, we have used the OHD to constrain the $\Lambda(t)$CDM as holographic and agegraphic dark energy models. However, the shortcoming of OHD is also
apparent because the amount of data points is too scarce to be comparable with other cosmic observations such as SNe Ia.
According to \cite{Hubble..parameter..MaCong}, by comparing the median figures of merit,
64 independent measurements of $H(z)$ will give a constraining results which can match the parameter constraining power of SNe Ia.
By applying the same method, we choose the standard $\Lambda$CDM model as the fiducial one to generate 64 $H(z)$ data points as a simulation
sample (In the rest we will use the abbreviation "SHD" for the simulated $H(z)$ data)
to constrain the $\Lambda(t)$CDM as holographic and agegraphic dark energy models. The simulated data sample is presented in figure \ref{fig:SHD}.

\begin{figure}[htb]
\begin{center}
\includegraphics[width=0.55\textwidth]{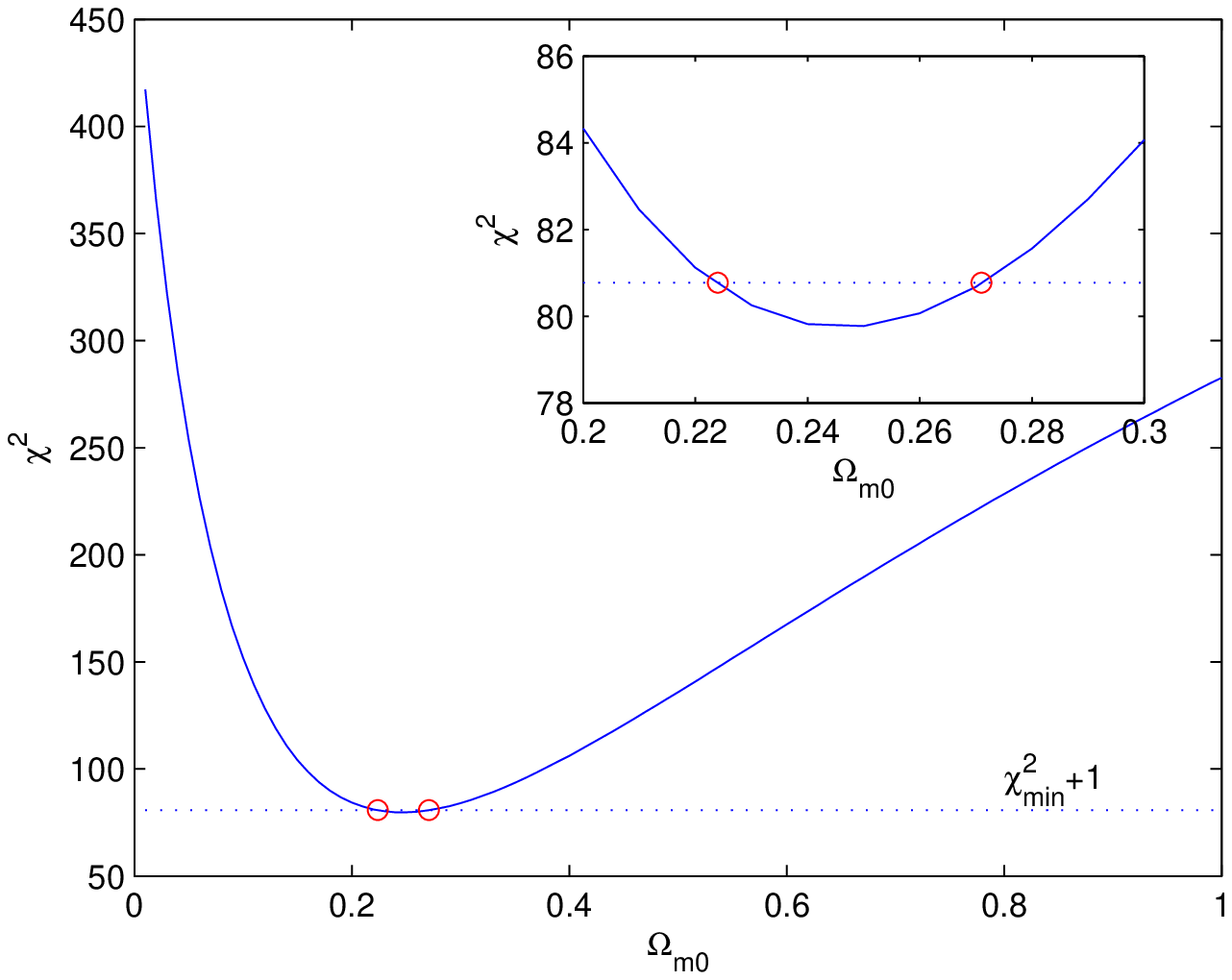}
\end{center}
\caption{The $\chi^{2}$ evolution with respect to $\Omega_{m0}$ for $\Lambda$CDM model by the use of SHD.
The circles stand for the $1\sigma$ confident interval and the insert gives zoom-in views around the 68.3\% confidence level.}
\label{fig:SHDLCDM}
\end{figure}

Using the SHD to constrain $\Lambda$CDM model, we get $\chi^{2}_{min}=79.7701$ and $\Omega_{m0}=0.25_{-0.03}^{+0.02}$ (figure \ref{fig:SHDLCDM}). This results is
consistent with the one under OHD and the $1\sigma$ interval is tighter. Thus it is natural to expect that with more data points,
the constraints on the holographic and agegraphic dark energy models will be more compact. The constraining results are summarized in Table.III

\begin{table}[!h]
\tabcolsep 0pt
\vspace*{-12pt}
\begin{center}
\def\temptablewidth{0.5\textwidth}
{\rule{\temptablewidth}{1pt}}
\begin{tabular*}{\temptablewidth}{@{\extracolsep{\fill}}cccc}
     Model  & $\chi^{2}_{min}$ &  $\Omega_{m0}$            & $c$                      \\   \hline
        HH  & 91.9684          &  $0.58_{-0.03}^{+0.02}$   & $0.65_{-0.03}^{+0.02}$   \\
       FEH  & 79.7261          &  $0.16_{-0.16}^{+0.22}$   & $0.65_{-0.15}^{+0.41}$   \\
       AU   & 79.7541          &  $0.25_{-0.04}^{+0.15}$   & $c\in(2.0,100.0)$        \\
       CT   & 79.7578          &  $0.25_{-0.03}^{+0.14}$   & $c\in(3.0,100.0)$
       \end{tabular*}
       {\rule{\temptablewidth}{1pt}}
       \end{center}
       \caption{constraints results by SHD}
       \end{table}

\begin{figure}[htb]
\begin{center}
$\begin{array}{cc}
\includegraphics[width=0.5\textwidth,height=0.4\textwidth]{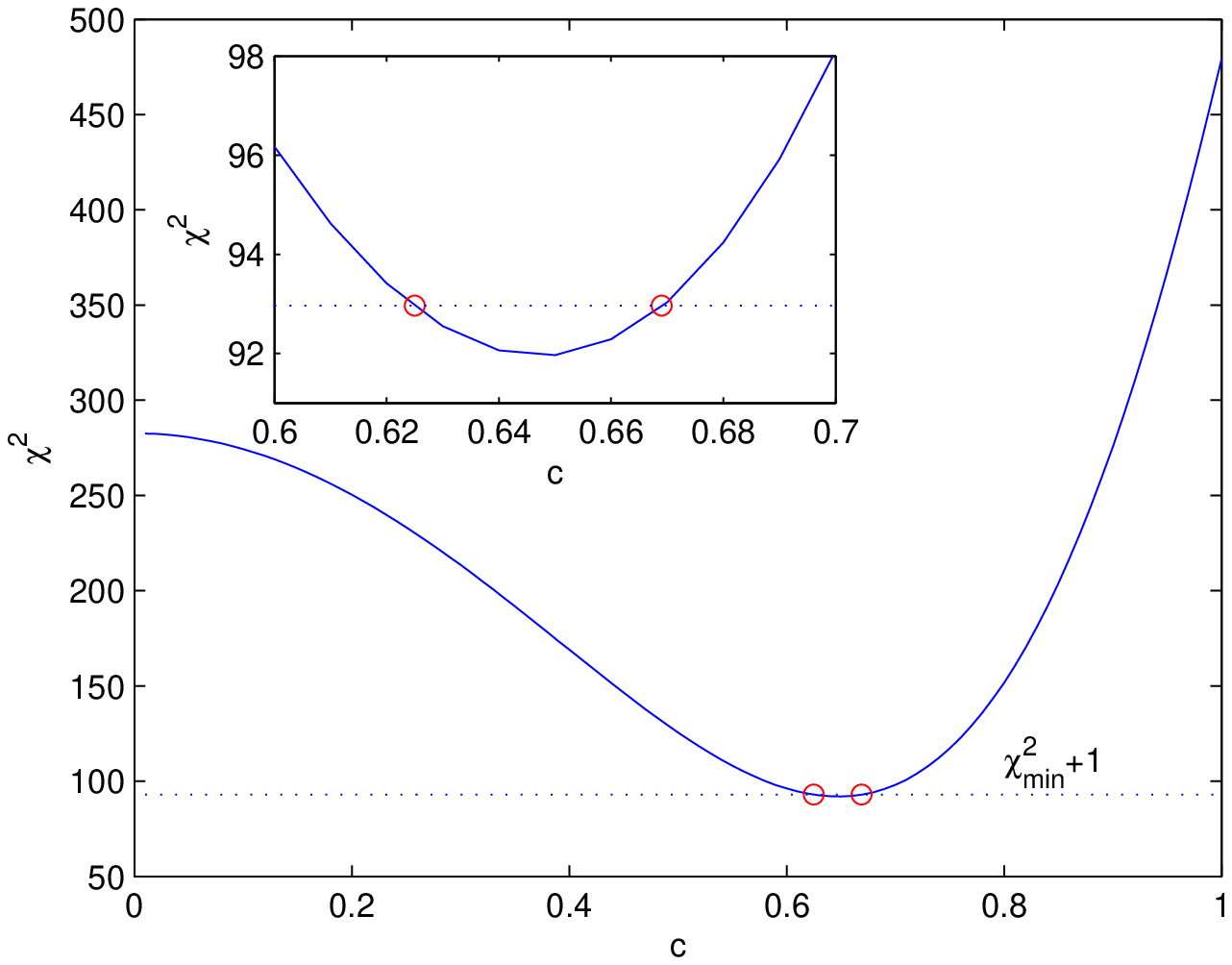} &
\includegraphics[width=0.5\textwidth,height=0.4\textwidth]{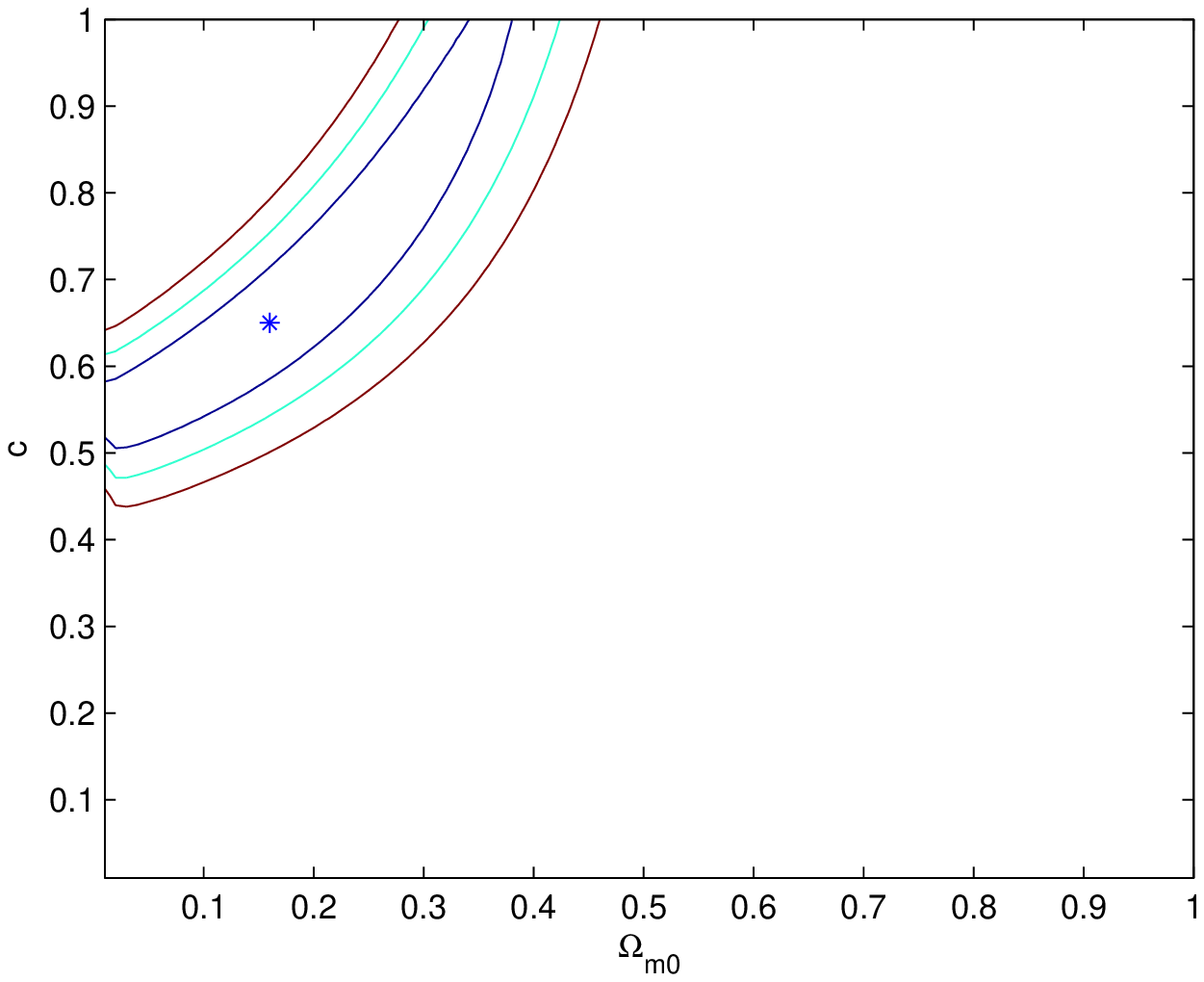}
\end{array}$
\end{center}
\caption{$Left$: The $\chi^{2}$ evolution with respect to $c$ for HH scenario constraining by SHD. The circles stand for the $1\sigma$ confident interval and the insert gives zoom-in views around the 68.3\% confidence level.
$Right$: The 68.3\%, 95.4\% and 99.7\% confident regions of FEH scenario from inner to outer by the use of SHD. The best-fit values is marked by the star.}
\label{fig:lengthSHD}
\end{figure}

\begin{figure}[htb]
\begin{center}
$\begin{array}{cc}
\includegraphics[width=0.5\textwidth,height=0.4\textwidth]{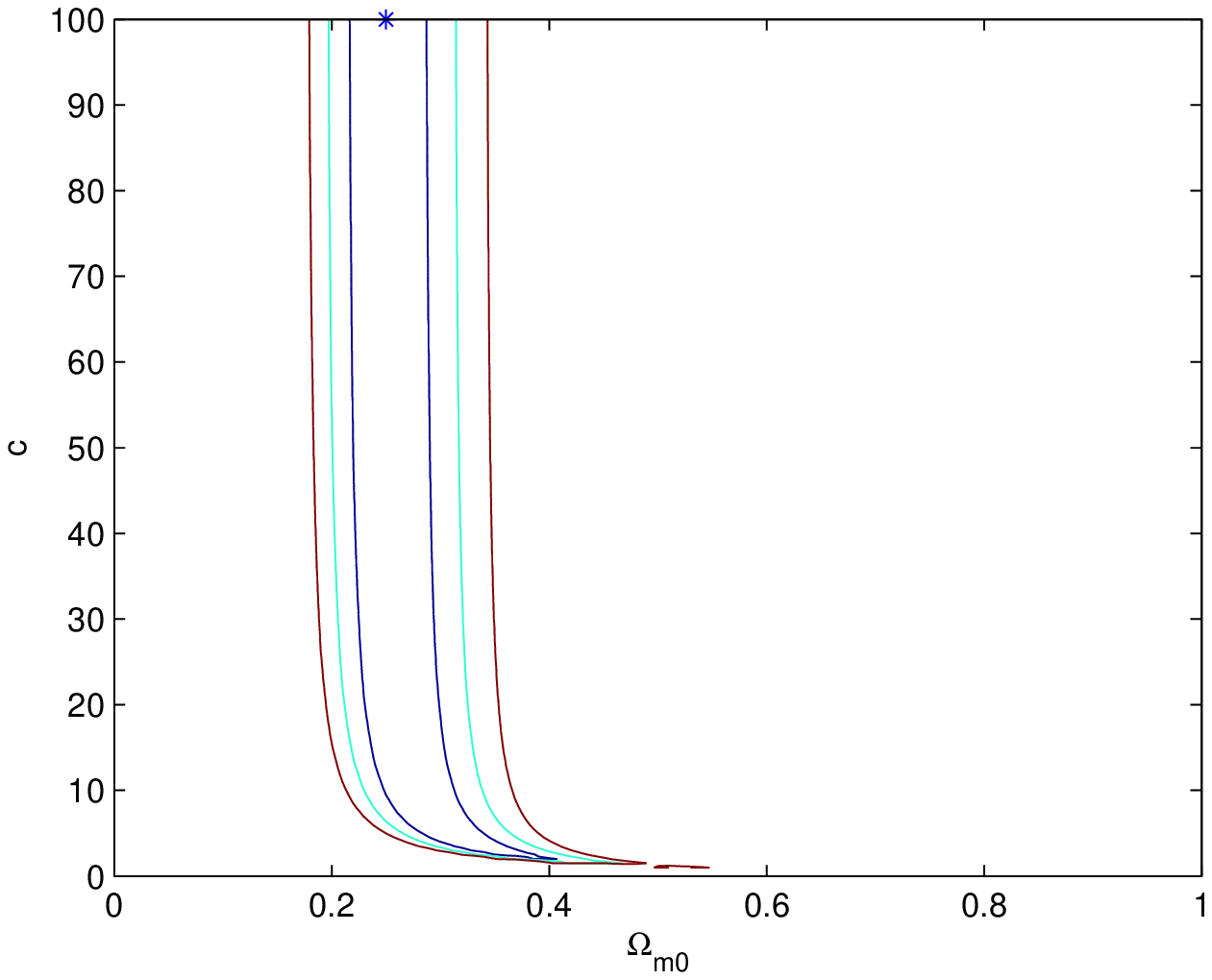} &
\includegraphics[width=0.5\textwidth,height=0.4\textwidth]{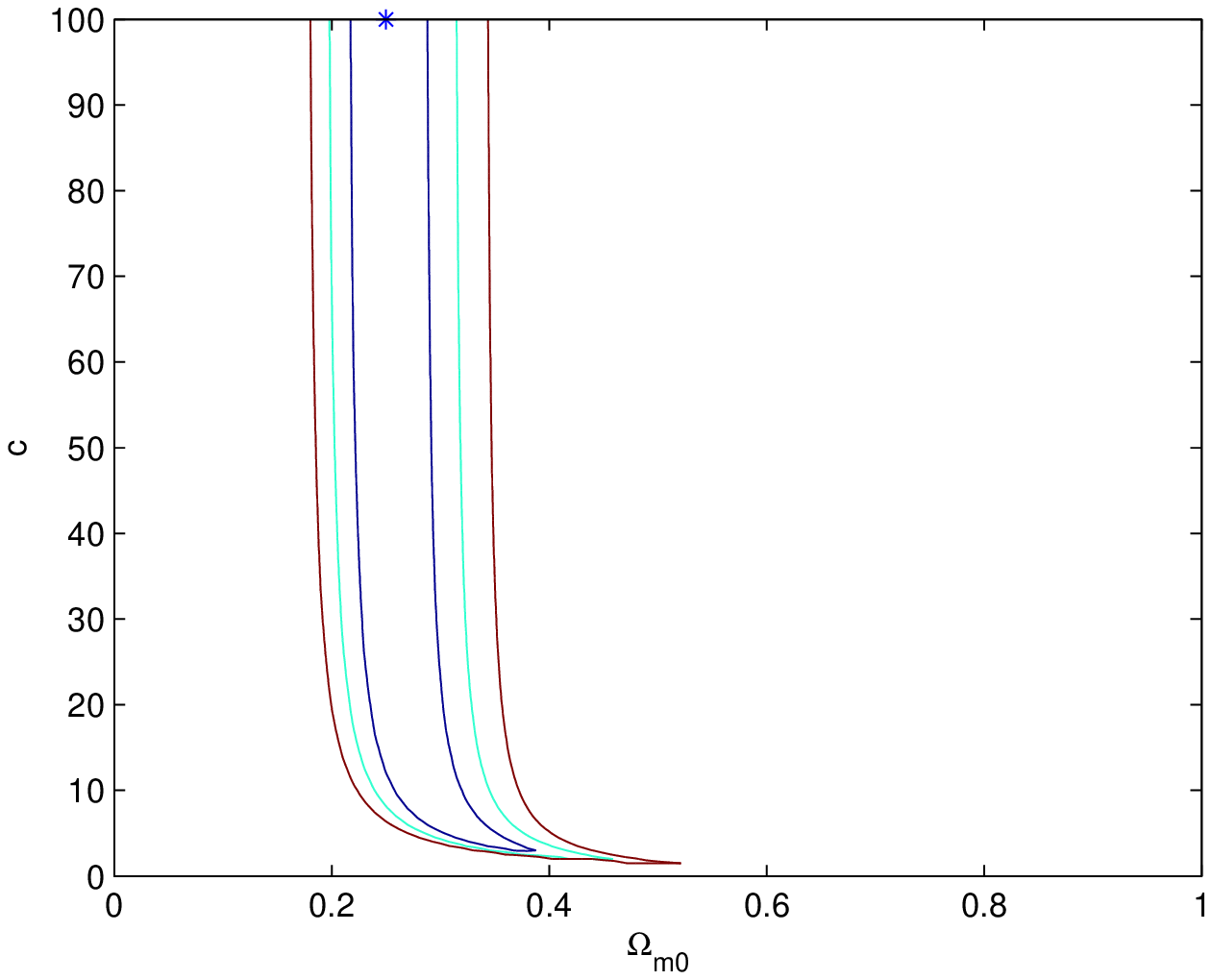}
\end{array}$
\end{center}
\caption{The 68.3\%, 95.4\% and 99.7\% confident regions from inner to outer in the $(\Omega_{m0}-c)$ plane by SHD constraints, while the star means the best-fit values.
$Left$: AU scenario; $Right$: CT scenario.}
\label{fig:timeSHD}
\end{figure}

In figure \ref{fig:lengthSHD} and \ref{fig:timeSHD}, we plot the confident regions for each models. Combined with Table.III,
we see that the constraint in HH scenario is more compact, however the results of large matter density and big $\chi^{2}_{min}$ value
show a great deviation from $\Lambda$CDM model and observations. For other three scenarios,
the more data points do not provide a much more tighter constraint of the $1\sigma$, $2\sigma$ and $3\sigma$
interval. However, the confident regions of these three models are tighter and narrower. This phenomena is consistent with \cite{Hubble..parameter..MaCong}. It implies: when defining
the FoM (Figure of Merit) as the area enclosed by the contour of possibility distribution at some confident level, the FoM will increase with enlarging
the data sample.
Additionally, similar to the OHD constraint, the FEH scenario also gives the smallest $\chi^{2}_{min}$.
Except that, the problem of lacking best-fit value and upper limit of $c$ still exists in the two time-scale models. This situation
implies that we need some other constraint conditions from the physical view.
Compared with the constraints using "Union 2 Compilation" of type Ia SNe\cite{lengthscale..timescale..Chen}, we see that all the 3 two-parameter models give consistent trends of confident regions.
And the best-fit values and confident interval of the undetermined parameters are also almost identical with the SNe Ia ones.

\section{Discussion and conclusion}\label{discussions}
We have explored the possibility of Hubble parameter as the cosmological constraint on the $\Lambda(t)$CDM as the holographic and
agegraphic dark energy models. As pointed out in \cite{Hubble..parameter..LinHui,fR..Hubble.vs.redshift..Carvalho}, the OHD play almost the same role as that of SNe Ia for the joint constraints on the $\Lambda$CDM
model. As a promotion of standard $\Lambda$CDM, the $\Lambda(t)$CDM may provide a wider range of adaptation of cosmological observations.
Recently, through the introduction of a length scale or time scale to give a $\Lambda(t)$ form has been investigated which is equivalent to the
holographic and agegraphic dark energy models. In order to track the deviation of these scenarios from the standard $\Lambda$CDM model, we use OHD
to constrain such $\Lambda(t)$CDM models and measure the deviations. We calculate the models introduced by the Hubble horizon (HH),
the future event horizon (FEH),
the age of universe (AU) and the conformal time (CT) since these scales can provide the present cosmic acceleration.

In order to avoid the dilemma that the amount of OHD is so scarce, we simulate a data sample of $H(z)$ (SHD) in the redshift range of $0.1<z<2.0$ by the use of
$\Lambda$CDM as the fiducial model. From the constraint results, we find that the enlarging of the data sample increases the parameter constraints apparently
for all four $\Lambda(t)$CDM models. Although the $1\sigma$,$2\sigma$ and $3\sigma$ confident intervals of the three two-parameters models do not
become narrower evidently, the areas of the confident regions in the parameter space shrink. This is expectable due to the bigger data sample.

However, in the constraints of two time scale scenarios, there are lacks of the best-fit value and upper limits of parameter $c$.
This is similar to the SNe Ia constraints \cite{lengthscale..timescale..Chen}. According to eq.(\ref{Eq:ageomega}) and (\ref{Eq:conformalomega}),
this phenomena reveals since $\Omega_{\Lambda}$ is insensitive to large value of $c$. Except that, when $c\rightarrow\infty$, the equation of state
of these two models will return to the standard $\Lambda$CDM model which $\Omega_{\Lambda}$ is a constant and favors a non-interaction term.
However, the lack of constraint of $c$ will derive $\rho_{\Lambda}$ to be a non-physical value. So we need some other constraint conditions from
the physical view in the future.

In order to measure the deviation of these $\Lambda(t)$CDM models from the standard $\Lambda$CDM model. We apply the $Om(z)$ and statefinder diagnostics and
the information criteria. These three methods can distinguish the dark energy models from the standard $\Lambda$CDM model on various aspects and focuses.
The calculation with the use of OHD shows quite consistent results of these methods. Choosing HH as the IR cut-off will derive a
quite different holographic dark energy model from the standard $\Lambda$CDM model and its relative derivation is apparent. As a another choice of length scale,
the FEH scenario shows better consistence with $\Lambda$CDM but still occupies prominent deviation. On the other hand, the two time models AU and CT scenarios
both possess very evident consistence from $\Lambda$CDM and can provide effective approximations.

From the above analyze and constraints, we find that our results from observational $H(z)$ data and simulated data are believable. The OHD
can not only provide sufficient constraints on the $\Lambda(t)$CDM models as holographic and agegraphic dark energy as the SNe Ia supply,
but also can measure the deviation of the cosmological models from standard $\Lambda$CDM model. Therefore, with more future high-$z$, high-accuracy
$H(z)$ determinations, OHD will give more and more important contributions in future cosmological researches.

\acknowledgments
This research is supported by the National Natural Science Foundation of China (Grant Nos. 10773002,
10875012), Scientific Research Foundation of Beijing Normal University under Grant No. 105116,
the National Science
Foundation of China (Grants No. 10473002, No. 11033005),
the Fundamental Research Funds for the Central Universities,
the Bairen program from the Chinese Academy of Sciences
and the National Basic Research Program of China grant No.
2010CB833000.



\end{document}